# Flow of gluten with tunable protein composition: from stress undershoot to stress overshoot and strain hardening

A. Louhichi[1*], M-H Morel[2], L. Ramos[1], A. Banc[1*]

[1] *Laboratoire Charles Coulomb (L2C), Univ. Montpellier, CNRS, Montpellier, France.*

[2] *UMR IATE, Université de Montpellier, CIRAD, INRAE, Montpellier SupAgro, 2 pl. Pierre Viala, 34070 Montpellier, France.*

*Corresponding authors: ameur.louhichi@umontpellier.fr and Amelie.banc@umontpellier.fr

**Abstract:**

Understanding the origin of the unique rheological properties of wheat gluten, the protein fraction of wheat grain, is crucial in bread-making processes and questions scientists since decades. Gluten is a complex mixture of two families of proteins, monomeric gliadins and polymeric glutenins. To better understand the respective role of the different classes of proteins in the supramolecular structure of gluten and its link to the material properties, we investigate here concentrated dispersions of gluten proteins in water with a fixed total protein concentration but variable composition in gliadin and glutenin. Linear viscoelasticity measurements show a gradual increase of the viscosity of the samples as the glutenin mass content increases from 7 to 66%. While the gliadin-rich samples are microphase-separated viscous fluids, homogeneous and transparent pre-gel and gels are obtained with the replacement of gliadin by glutenin. To unravel the flow properties of the gluten samples, we perform shear start-up experiments at different shear-rates. In accordance with the linear viscoelastic signature, three classes of behaviour are evidenced depending on the protein composition. As samples get depleted in gliadin and enriched in glutenin, distinctive features are measured: (i) viscosity undershoot suggesting droplet elongation for microphase-separated dispersions, (ii) stress overshoot and partial structural relaxation for near-critical pre-gels, and (iii) strain hardening and flow instabilities of gels. We discuss the experimental results by analogy with the behaviour of model systems, including viscoelastic emulsions, branched polymer melts and critical gels, and provide a consistent physical picture of the supramolecular features of the three classes of protein dispersions.

# Introduction:

Gluten extracted from wheat flour is one of the most important commercial plant proteins isolate. It is used to improve the baking quality of cereal products consumed daily by many human beings through the consumption of bread, pasta, and biscuits among others [1, 2]. The exceptional mechanical properties of the gluten network, is essential for the growth of bubbles in bread dough and renders gluten very attractive as a food texture additive. Mechanical properties of gluten have been investigated for a long time, but gluten complexity in terms of composition and solubility makes studies difficult [3]. The gluten network comprises two main classes of proteins; the polymeric glutenins that have the ability to form inter- and intra-disulfide bonds and the monomeric gliadins that participate in the network through hydrogen bonding exclusively [4, 5]. Monomeric gliadins can be isolated from gluten extracted from wheat, whereas glutenin-rich extracts are never devoid of gliadins. Glutenin polymers display a very wide range of molecular weight $M_w$ (from 100 to 1000 kg/mol) [1], and are considered as being soluble in acidic solutions. Gliadins on the other hand have $M_w$ comprised between 25 and 60 kg/mol and are soluble in aqueous ethanol (with an ethanol volume fraction between 50 and 70%). All gluten proteins are considered as being water-insoluble at neutral pH.

Nevertheless, protein mixtures comprising gliadins and glutenins dispersed in pure water at a protein concentration above 30% w/w form homogeneous viscoelastic networks. Unveiling the peculiar contribution of each of the main classes of gluten proteins to the mechanical properties of gluten network and wheat dough in relation to their structural features is of a great importance but scientifically challenging, especially because of gluten polymorphism and polydispersity, and solubility issues [6].

Different protocols have been used in previous studies to characterize the rheology of gluten gels [7-13]. Of particular interest, Ng et al. [12, 13] investigated a native gluten dough containing 63% of water by weight through different protocols in the linear and nonlinear regime (including small and large amplitude oscillatory shear, step strain relaxation, creep, start-up of steady shear and uniaxial flow). The originality of this work consisted in the successful modification of the well-known power law relaxation model of critical gels [14] to describe the linear and nonlinear viscoelasticity responses of the gluten gel when investigated using several rheological protocols. Strain softening of gluten was measured by creep when the samples were submitted to sufficiently large stresses [10], and by large amplitude oscillatory shear (LAOS) tests at moderate strains. Strain softening was modelled using a non-linear network destruction term that reflects the reduction in network connectivity as proteins are increasingly stretched. In addition, at larger strains (of the order of $\gamma=6$), strain hardening was evidenced before sample rupture due to the finite extensibility of the network [13]. Although evidencing some of the remarkable mechanical properties of gluten, previous studies fail to provide a complete mechanistic understanding of the rheology of gluten, in part because they remain limited in terms of samples and lack a proper control of the protein composition and biochemistry, and of the resultant microstructure. To overcome these limitations, we have developed protocols to produce model gluten extracts with a tunable proportion of gliadins and glutenins [15]. We first focused on the structure and the linear viscoelasticity response of model glutens comprising comparable amounts of gliadin and glutenin, as in native gluten, in a blend of water and ethanol [16-19]. In a 50/50 v/v water/ethanol mixture, homogenous transparent samples are obtained for a large range of protein composition and concentration [19, 20]. The samples display a disordered polymeric structure and their linear viscoelasticity obeys the framework of the near critical gel theory [21]. For near critical gels, the self-similarity of the clusters that eventually percolate at a critical point results in criticality in the linear viscoelastic response [22, 23]: the complex moduli and the relaxation times vary as a power law of the frequency, with a critical exponent related to the fractal dimension of the stress bearing network [24]. As consequence, following the near critical gel theory, we have demonstrated that the linear rheology of model gluten gels with comparable amounts of gliadin and glutenin but different protein concentrations and sample ages can be described by a unique master curve by applying a time-cure-concentration superposition principle [14, 17, 22, 23, 25, 26]. The spontaneous gelation of samples was attributed to the rearrangement of both intermolecular hydrogen and disulphide bonds [17]. More recently, we have successfully extended the time-cure-concentration principle to include solvent quality using water/ethanol mixtures with variable compositions from pure water to 60% v/v ethanol [27]. An important conclusion of this study is that the general framework to rationalize the structural and mechanical properties of model gluten protein extracts dispersed in a water/ethanol mixture, a good solvent for gluten proteins, also holds with pure water, which is usually considered as a bad solvent for gluten, thus extending our investigations towards food applications.

As for the impact of protein composition on rheology, it is generally accepted that gluten viscosity is related to gliadins, whereas elasticity and stiffness are associated to glutenins [3, 28]. Crude fractions of gliadin and glutenin were investigated in the linear regime but using a denaturing solvent (3M urea) that significantly modifies interactions, which are crucial for the gluten network [29]. Recently, Large amplitude oscillatory shear (LAOS) experiments performed on these fractions once dispersed in water showed that the nonlinear response of gliadin samples is essentially viscous and frequency-dependent, whereas the glutenin-rich fraction displays a stiffer response independent of frequency, thus confirming the overall physical understanding of the role of two main classes of proteins in gluten [30]. To better unveil the role of the two classes of proteins and their interplay in the rheology of gluten, we have studied the effect of gluten compositions on the structure and linear viscoelasticity of gluten gels using water/ethanol 50/50 %v/v as a solvent [20]. Thanks to an asymmetrical flow field flow fractionation technique, we have shown that dilute solutions are mainly composed of monomeric and polymeric species when the glutenin content is low, whereas additional supramolecular objects of hundred nanometers, namely assemblies, are identified in increasing proportion when the samples get enriched in glutenin (mass fraction of glutenin>25% ) [31]. Interestingly, these assemblies are composed of both high molecular weight glutenin polymers and gliadins [18, 31], demonstrating interactions between gliadins and glutenins. Moreover, the emergence of these assemblies in dilute regime coincides with the emergence of the viscoelasticity in semi-dilute samples of equivalent protein composition [20]. Assemblies can thus be identified as playing a key role in the onset of gelation, and might be considered as precursors of the self-similar clusters in the framework of near critical gel model. Following these studies, this manuscript aims at exploring the nonlinear viscoelastic properties of model gluten samples with controlled composition, using the start-up shear protocol.

Shear start-up protocol is classically used to monitor the transient response of different model systems, such as polymer melts and solutions [32], colloidal gels and glasses [33], worm-like micelles solutions [34] and emulsions [35]. A common phenomenology emerges from these studies. At low shear rates, a monotonic increase of the stress with time goes towards a plateau value, corresponding to a steady state. At higher shear rates or by varying the molecular characteristics of the system (concentration, volume fraction, molecular weight, particle size, etc ...), the stress passes through an overshoot, before reaching its steady state. The general physics behind the transient behaviour consists in a competition between the shear rate and the relevant relaxation rate in the system. When the shear rate is slower compared to the sample characteristic relaxation rate, no overshoot is observed because the structure has time to relax the stress within the shear time. However, in the regime where the shear rate is higher than the sample relaxation rate, a stress overshoot is measured. The position and the amplitude of the overshoot depends on the maximum deformation allowed by the structure and its relaxation at a given shear rate. The steady state regime on the other hand reflects the competition between the stress induced by the flow and the stress relaxed by the sample through structural changes. Despite the fact that the general physics that drives the overshoot is the same, the structural origin differs from a system to another, and questioning the physical origin of stress overshoots is still gathering a lot of attention [36]. In addition, for some systems and in specific conditions (such as high shear rate, confinement…), a stress undershoot was observed; however its origin remain not totally understood so far [35, 37-40].

In the present paper, we provide a systematic investigation of model gluten samples prepared in water at a fixed protein concentration but with different compositions in terms of glutenin content. We use the shear start-up protocol that allows one to study the samples with different shear rates and, in the same time, reach a high cumulative deformation up to 2000. The manuscript is organised as follows. We first describe the materials and the techniques. We then present the experimental data in the linear and in the non-linear regimes that evidence three classes of rheological behaviours depending on the protein composition. We finally discuss results in comparison with different model systems to propose a structural view of samples.

## Materials and Methods:

## Materials:

We investigate gluten samples with a fixed total protein concentration but different proportions of monomeric gliadin and polymeric glutenin. Six samples with mass fraction of glutenin (GLU) ranging from 7 to 66% are investigated. The extraction of the gluten powder with controlled and tuneable compositions is detailed elsewhere [15]. In brief, a native gluten solution in a water/ethanol 50/50 %v/v is stirred for 19 h at 20 °C and then centrifuged (30 min, 15,000g). The resulting supernatant is then quenched for 1 h at a low temperature, $T_q$, to yield a liquid−liquid phase separation into a light phase and a dense phase. Each phase contains different amounts of gliadin and glutenin depending on the quenching temperature. Both phases are frozen at −40 °C, and then freeze-dried and ground. The compositions of the resulting powders are probed by chromatography [15]. The dense phases provide gluten extracts with GLU=50% ($T_q$=2°C), GLU=52% ($T_q$=3°C), GLU=57% ($T_q$=6°C) and GLU=66% ($T_q$=9°C), and the light phases provide gluten extracts with GLU=7% ($T_q$=3 °C) and GLU=23% ($T_q$=9 °C).

Samples are prepared by dispersing the required mass of gluten extract in the appropriate volume of deionized pure water to obtain a fixed concentration of 500 g/L for all samples using a specific volume value of gluten proteins, $v = 0.76$ mL/g. The deionized water contains 0.1% g/g of sodium azide ($NaN_3$) to prevent microbial growth. Dispersion of the protein extracts in the solvent is achieved through hand-mixing using a stainless-steel spatula for 3 min. The dispersions are then kept to rest for 5 days at room temperature, prior to measurements, in order to obtain homogenous samples. The relatively long rest period allows an efficient hydration of the proteins and a complete relaxation of the stresses induced by the mixing step. Transparent samples are obtained for GLU≥50%, while turbid samples, macroscopically stable over several weeks, are obtained for GLU=7% and 23% (Fig. 1). Light microscopy imaging of the turbid samples reveals the presence of micrometric spherical objects, suggesting a liquid-liquid phase separation between a phase enriched in protein and one depleted in proteins. This observation is consistent with a previous study which indicate that gliadins become insoluble in distilled water for concentration higher than 15% wt [41].

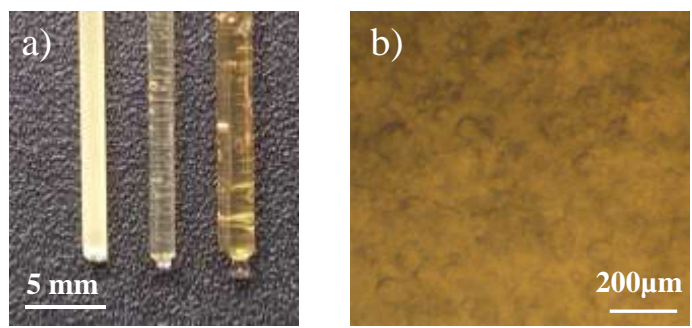

*Figure 1: a) Pictures of samples with GLU=7, 52, and 66% (from left to right) inserted in thin capillaries. b) Light microscopy image of the sample with GLU=7%.*

## Methods:

Linear and nonlinear rheology of the gluten samples is measured using an MCR 302 rheometer (Anton Paar, Austria), operated in strain-controlled mode. All experiments are performed at a temperature of 25 ºC achieved by means of a Peltier element with a precision of ± 0.2 ºC. We use cone-plate geometries, with different diameters (8, 25 or 50 mm), depending on the sample viscoelasticity. The gap between the cone and the plate is set to its predefined value (101 μm, 53 μm and 51 μm for the cone with diameter 50 mm, 8 mm and 25 mm, respectively). The sample edge and the upper cone are immersed in a bath of silicon oil to avoid solvent evaporation. To minimize sample slip, a rough bottom plate is systematically used. The cones with diameter 8 mm and 25 mm are rough, and the one with dimeter 50 mm is smooth. The roughness of the rough tools is ~6 μm in height.

The linear viscoelasticity (LVE) of all samples is measured through dynamic frequency sweeps in the linear regime defined by means of independently performed dynamic strain sweeps. The linear regime is defined as the range of deformation where the storage (G') and the loss (G") moduli are constant with the strain amplitude $\gamma_0$. All the samples stayed in the linear regime up to a strain $\gamma_0=0.1$. LVE spectra are collected over 4 decades of frequency, from 100 rad/s to 0.01 rad/s.

The nonlinear viscoelasticity (NLVE) is measured through a so-called shear start-up protocol [42, 43] which consists in applying a fixed steady shear rate, $\dot{\gamma}$, for a certain time, t, and monitoring the time evolution of the transient stress $\sigma^+$. We impose different shear rates, $\dot{\gamma}$, from $0.1\ s^{-1}$ to $10\ s^{-1}$, and the duration of each experiment at a given shear rate is tuned until a constant value for the stress $\sigma^+$ is reached. For each sample, a series of successive measurements run at growing shear rates, in the range $(0.1 - 10)\ s^{-1}$ are performed with a relaxation period of 2400 s at $\dot{\gamma} = 0$ between each different shear start-up. The series are bordered by LVE tests to check the reproducibility of the LVE spectra after the nonlinear deformation.

## Results:

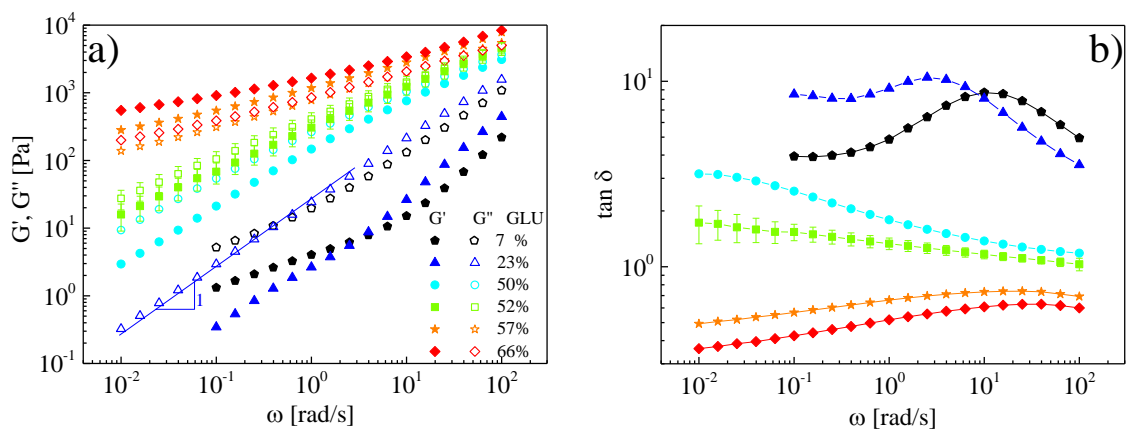

*Figure 2: Linear viscoelastic data of gluten samples with different protein compositions as indicated in the legend as a function of frequency: a) storage G' (filled symbols) and loss G'' (empty symbols) moduli, the line has a slope of 1; b) tan δ=G''/G'.*

Figure 2 depicts the linear viscoelasticity for samples with different protein compositions. The frequency dependence of the storage, G', and loss, G", moduli is shown in Figure 2a and tanδ=G"/G' is plotted in Figure 2b. The error bars on the data for the sample with GLU=52% represent the standard deviation as computed based on a dozen of measurements performed on independent samples prepared and measured in the same conditions. The relative error on G' and G" ranges between 30 % and 40 % and the relative error on tan δ is lower (8 to 20%). In addition, note here that the data of G' for the two samples depleted in glutenin (GLU=7 and 23%) are not plotted at low frequencies because measurements are not reliable (weak signal due to the rheometer torque limits). We observe that the samples present qualitative and quantitative different features as their protein composition varies. First, the higher the content in glutenin the stronger the viscoelasticity of the gluten sample is. The strengthening of the viscoelasticity is evidenced by the fact that the shear viscoelastic moduli, G' and G'', systematically increase when GLU increases (Fig. 2a). The frequency dependence of the LVE spectra also changes significantly with the protein composition. On the one hand, for the two samples with the lowest amounts of glutenin (GLU=7% and GLU=23%), the loss modulus, G'', is significantly higher than the storage modulus, G'. The loss modulus exhibits a power law dependence with the frequency with an exponent close to 1 (G''~ ω) at low frequency, which is the signature of a completely flowing fluid. The evolution of the elastic modulus, G', is more complex with a faster increase with frequency at high frequency and a lower one at low frequency, with a cross-over between the two regimes around 10 rad/s, and 2 rad/s, for the sample with GLU=7%, and GLU=23%, respectively. The viscoelastic spectra of the two samples rich in gliadin are similar to those observed for blends of immiscible polymers or emulsion of viscoelastic fluids [44, 45], in agreement with the sample structure. On the other hand, for the two samples with intermediate amounts of glutenin (GLU=50 and 52%), one also measures a loss modulus larger than the storage modulus over the whole experimentally accessible frequency window, hence indicating that these two samples are fluid. By contrast, for the two samples comprising the highest amounts of glutenin (GLU=57 and 66%), G' is larger than G'' over the whole frequency window with the emergence of an elastic plateau at low frequency. Overall, the change of the behaviour of G' and G'' and the occurrence of a plateau for the storage modulus confirms the transition from a liquid-like behaviour (G''>G') to a solid-like one (G'>G'') by increasing the proportion of glutenin in the protein extract. Accordingly, tan δ decreases with frequency for homogeneous liquid-like samples (G''>G') and its frequency dependence is weaker by increasing GLU, as shown in Figure 2b. For solid-like samples (G'>G"), tan δ is smaller than 1 and increases with frequency. For the microphase-separated samples, by contrast, tan δ displays a non-monotonic evolution with a minimum at low frequency presumably related to the characteristic relaxation time of droplets (of the order of few seconds) and a maximum at higher frequency that depends on the viscosity contrast between the two phases and the interfacial tension between the two phases [44].

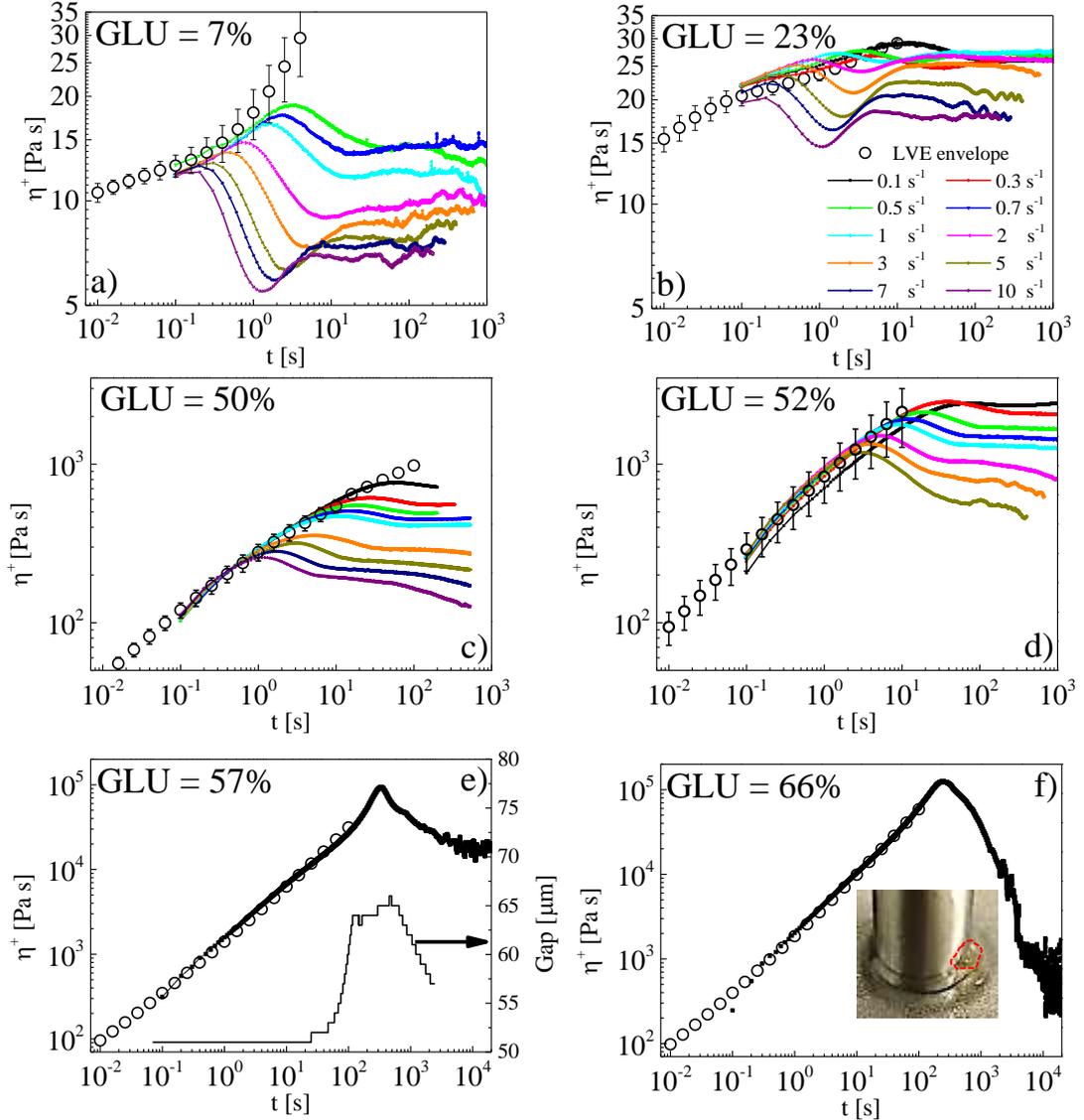

*Figure 3: Transient shear viscosity η+ as a function of time, t, for gluten samples with different protein compositions (from GLU=7% in (a) to GLU=66% in (f) as indicated in the left corner of each graph) and different shear rates, as indicated in the legend (colored lines), and linear viscoelastic envelopes (open black circles). In (e), the black thin line shows the evolution of the gap between cone and plate during the experiment at $\dot{\gamma} = 0.1\ s^{-1}$. In (f), the picture in the inset shows that part of the sample is expelled out of the gap at time t=100 s. The scale of the picture is given by the diameter of the upper cone (8 mm).*

We compute the transient viscosity, $\eta^+ = \frac{\sigma^+}{\dot{\gamma}}$, where $\dot{\gamma}$ is the imposed shear rate, and $\sigma^+$ is the measured stress, as a function of time, t, where the origin of time is the time at which the shear rate is applied. Figure 3 shows the time evolution of $\eta^+$ at different shear rates, in the range (0.1- 10 $s^{-1}$) for samples prepared with gluten with different compositions, as indicated in the legends. Note that for the samples with the lowest amount of glutenin (GLU=7%), we show data sets for shear rates $\geq 0.7\ s^{-1}$ because data at lower shear rates are poorly reliable due to the very weak measured torque. On the other hand, for the samples with the two largest amounts of glutenin (GLU=57 and 66%), we plot only the transient viscosity obtained during the first shear start-up experiments ($\dot{\gamma} = 0.1\ s^{-1}$) because severe instabilities occur due to transducer overloading (gap opening and sample expulsion from the geometry gap). We also compute for all samples the linear viscoelastic envelope. The envelope is calculated through a direct transformation of the dynamic linear data by applying the Cox-Merz rule $\eta(\dot{\gamma}) = \eta^*(\omega)|_{\omega=\dot{\gamma}}$ [46] in conjunction with the Gleissle relationship $\eta^+(t) = \eta(\dot{\gamma})|_{\dot{\gamma}=1/t}$ [47] using the complex viscosity, $\eta^* = \frac{\sqrt{G'^2+G''^2}}{\omega}$. In Figure 3, the linear envelope is plotted in open black circles and corresponds to the mean complex viscosity, $\eta^*$, obtained from the LVE measured before and after the nonlinear experiment series (for samples with GLU=7, 23, 50 and 52%). The error bars represent the standard deviation. The relative error ( ~ 40%) iscomparable to the relative error computed from the repetition of the LVE measurements (as described above, see Fig. 2a) and is presumably due to a slight sample ageing. For samples with GLU=57 and 66%, the linear envelope is obtained from LVE data measured before the NLVE protocol, because the NLVE protocol significantly modifies the LVE signature of these samples. Interestingly, for all samples, we find consistency between the complex viscosity and the transient viscosity at short times.

From the transient viscosity and its evolution with the imposed shear rates three classes of samples can be evidenced, based on their amount of glutenin: (i) for the two samples most depleted in glutenin (GLU=7 and 23%), $\eta^+(t)$ exhibits a noticeable undershoot, before reaching a steady state regime, at large shear rates ($\dot{\gamma} \geq 1\ s^{-1}$ for GLU=7% and $\dot{\gamma} \geq 0.3\ s^{-1}$ for GLU=23%); (ii) For the two samples with an intermediate amount of glutenin (GLU=50 and 52%), a stress overshoot is measured: the transient viscosity exhibits a maximum before reaching a lower steady state. (iii) The two samples rich in glutenin (GLU=57 and 66%) are gels. They exhibit flow instabilities during shear start up experiments. A strain hardening is measured for the sample with GLU=57% that is accompanied by an increase of the gap (thin black line in Fig. 3e) presumably because of the growing high normal force exceeding the rheometer limit. The gap opening alters the result, and suggests that the hardening could be stronger that the one measured. In the same vein, no clear hardening is observed for the sample with GLU=66%, presumably because of the sample being expelled out of the gap before the potential development of a significant hardening (see picture in the inset of Fig. 3f).

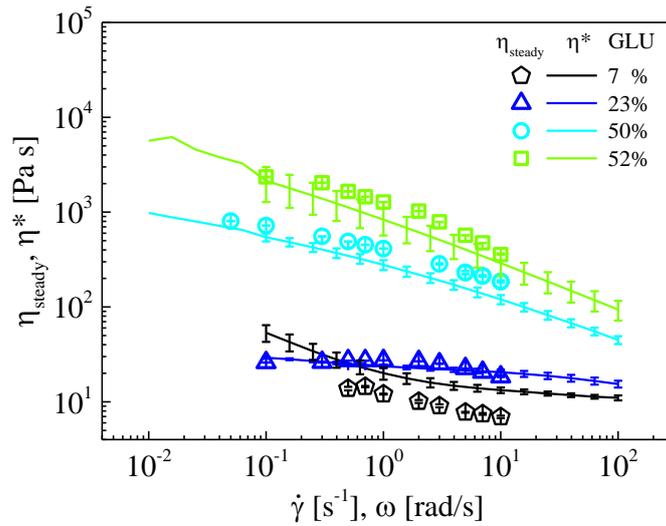

*Figure 4: Steady viscosity $\eta_{steady}$ (open symbols) as a function of the shear rate $\dot{\gamma}$, and complex viscosity $\eta^*$ (thick lines) as a function of the frequency, $\omega$, for gluten samples with different protein compositions as indicated in the legend.*

At long time, a steady viscosity, $\eta_{steady}$, is reached for samples with GLU=7, 23, 50 and 52%, although some irregularities (slight decrease at high shear rate or some erratic oscillations) are measured. In that case the steady viscosity value reported in figure 4 is averaged over the most stable range. Figure 4 shows the steady viscosity $\eta_{steady}$, as a function of the shear rate, $\dot{\gamma}$, together with the complex viscosity $\eta^*$ as a function of the frequency, $\omega$. Both quantities roughly superimpose within the experimental errors, for all viscous samples (GLU≤52%), except for the more fluid sample (GLU=7%) where η* is systematically slightly larger than $\eta_{steady}$. The fair collapse of both viscosities suggests that the nonlinear flow does not induce any irreversible damage in the samples investigated. The complex viscosity for the two samples with the lowest proportion of glutenin (GLU=7% and 23%) shows a very weak decrease with the shear rate. For samples with intermediate amount of glutenin (GLU=50 and 52%), both viscosities exhibit a clear shear thinning behaviour.

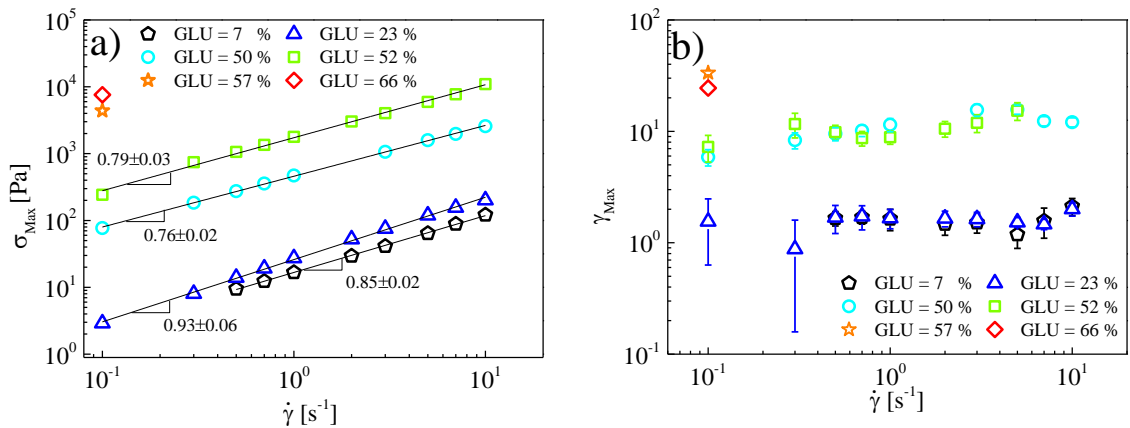

*Figure 5*: *Maximum stress, $\sigma_{Max}$ (a), and maximum strain, $\gamma_{Max}$ (b), as a function of the shear rate $\dot{\gamma}$ for gluten samples with different protein compositions as indicated in the legend. In (b) error bars account for the broadness of the viscosity peak.*

A simple analysis common to all the shear start up experiments performed with the six different samples is to collect for each data set, the maximum stress, $\sigma_{Max}$, and the strain at which it is reached, $\gamma_{Max}$. The shear rate evolutions of $\sigma_{Max}$ and $\gamma_{Max}$, are plotted in Figure 5a and Figure 5b, respectively. We observe that $\sigma_{Max}$ varies as a power law with the shear rate for all samples investigated. Best fits of the experimental data yield comparable exponents of the power law exponent for all samples (0.8±0.1), although slightly larger for the samples depleted in glutenin. Overall, the evolution of $\sigma_{Max}$ with the amount of glutenin in the sample (from $\sigma_{Max} = 3$ Pa to $\sigma_{Max} = 7564$ Pa at $\dot{\gamma} = 0.1$ s$^{-1}$ as GLU increases from 7% to 66%) reflects the sample strengthening with the amount of glutenin in the protein mixture. On the other hand, we measure that for all fluid samples, $\gamma_{Max}$ does not depend on the shear rates, but have markedly different values for the samples showing a stress undershoot ($\gamma_{Max} \cong 1.6 \pm 0.3$ for samples with GLU=7 and 23%), as compared to the samples showing a stress overshoot ($\gamma_{Max} \cong 10 \pm 4$ for samples with GLU=50 and 52%). The different values for the two types of samples suggest different physical processes, as will be discussed below. For the gel samples, $\gamma_{Max}$ could only be measured at $\dot{\gamma} = 0.1$ s$^{-1}$, and corresponds to the strain above which instabilities and eventually sample damages occur. The numerical values are much higher than for the fluid samples ($\gamma_{Max} \cong 34$ for GLU=57% and $\gamma_{Max} \cong 25$ for GLU=66%). In the following, we analyse separately the behaviour of the fluid samples showing stress undershoot and stress overshoot.

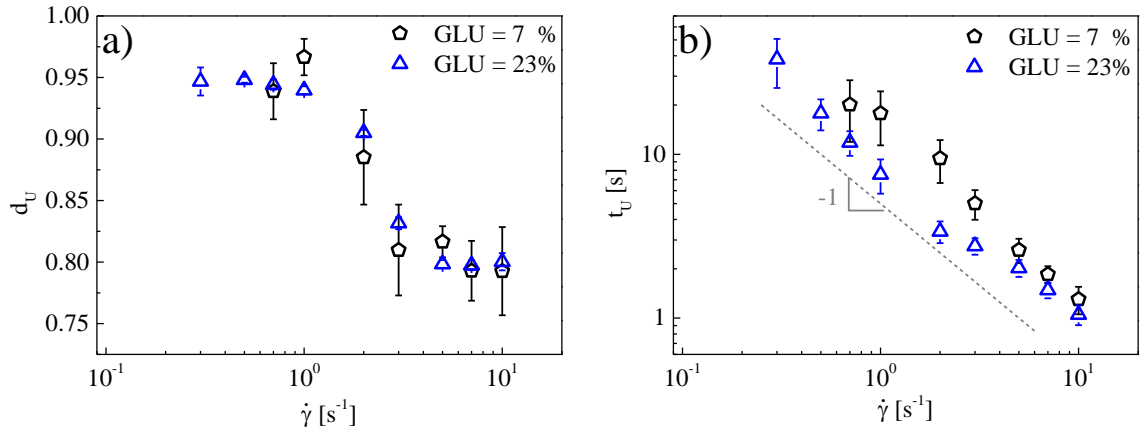

*Figure 6: The undershoot characteristics: (a) the amplitude $d_U = \frac{\sigma_{Min}}{\sigma_{Steady}}$, (b) the time of its occurrence, $t_U$, as a function of the shear rate for the samples with GLU=7% and 23% as indicated in the legend. The doted gray line in (b) has a slope of -1.*

The two samples with the lowest amounts of glutenin (GLU=7% and 23%) show a stress undershoot before reaching a steady state (Fig. 3a,b). We define the amplitude of the stress undershoot as $d_U = \frac{\sigma_{Min}}{\sigma_{Steady}}$, with $\sigma_{Min}$ and $\sigma_{Steady}$, the minimum stress, and the stress in the steady state at long time, respectively. We find that the data sets for the two samples superimpose nicely and that $d_U$ decreases continuously as the shear rate increases, from 1 down to 0.8 (Fig. 6a). The time of occurrence of the undershoot, $t_U$, defined as the time corresponding to the minimum stress $\sigma_{Min}$, is plotted in Figure 6b as a function of the shear rate $\dot{\gamma}$. For both samples, $t_U$ is measured to be inversely proportional to the shear rate, indicating an occurrence of the undershoot at a constant strain of $\gamma_U = 15 \pm 3$ for GLU=7% and $\gamma_U = 9.3 \pm 1.6$ for GLU=23%.

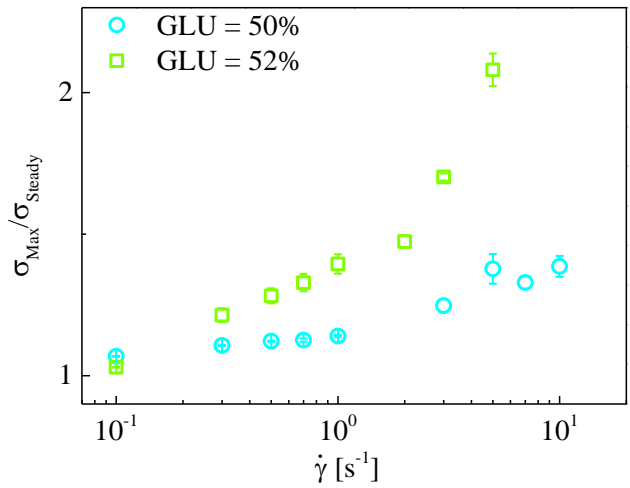

*Figure 7: Maximum stress normalized by the steady state stress $\sigma_{Max}/\sigma_{Steady}$ as a function of the shear rate $\dot{\gamma}$ for gluten samples with GLU=50 and 52%, as indicated in the legend.*

The two samples comprising intermediate amounts of glutenin show a stress overshoot (Fig. 3c,d). The amplitude of the overshoot, $\sigma_{Max}/\sigma_{Steady}$, with $\sigma_{Steady}$ the stress in the steady state, is plotted as a function of the shear rate in Figure 7 for the two samples with GLU=50 and 52%. In the two cases, we find a weak increase of $\sigma_{Max}/\sigma_{Steady}$ with the shear rate (at most by a factor of 2 when the shear rate varies by two orders of magnitude). Nevertheless, significantly different results are obtained for the two samples, with a stress overshoot systematically higher for the sample with the largest amount of glutenin and a stronger dependence with the shear rate.

## Discussion:

We have identified three families of samples based on their distinctive linear and non-linear viscoelasticity features. Below we discuss sequentially the three families, whose structures are schematically sketched in Figure 7:

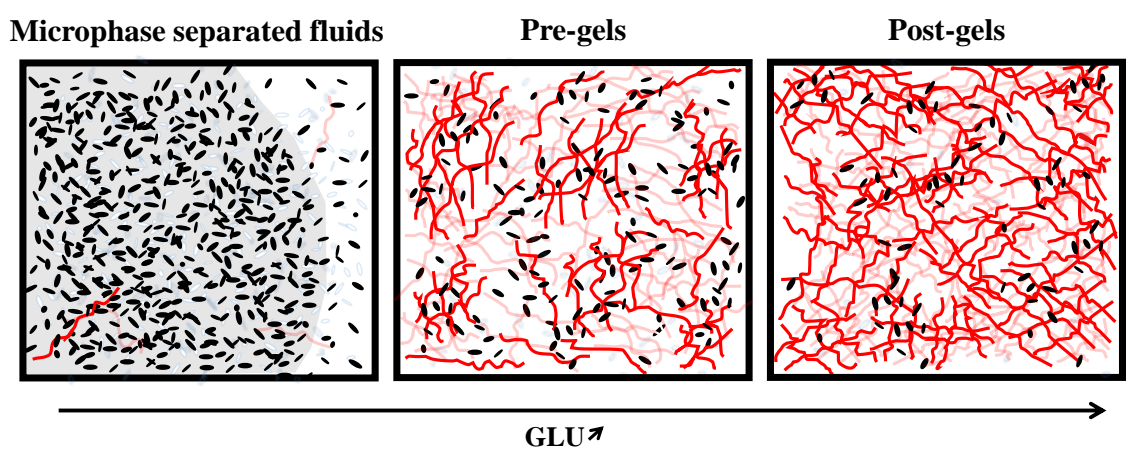

Figure 8. *Schematic illustration for the structure of the different families of samples investigated. The black ellipsoidal dots represent gliadins and the red lines represent glutenin polymers.*

The samples rich in gliadins (GLU ≤ 23%) form microphase-separated dispersions in water and are the ones with the weakest viscoelasticity. They are flowing systems with a linear viscoelastic signature that can be compared to that of viscoelastic emulsions. Their viscoelasticity is dominated by the loss modulus that is proportional to the frequency, ω, while the storage modulus shows a nearly ω² dependence at high frequency with a transition towards a weaker dependence at lower frequency. It suggests the presence of contrasted relaxation times in the sample that can be attributed at high frequency to the viscoelastic continuous phase and at low frequency to the geometrical relaxation of droplets according to the model developed by Palierne and collaborators for viscoelastic emulsions [44]. The lower moduli measured at high frequency for the sample with GLU=7% suggests that the terminal relaxation of the continuous phase is shorter than that for the sample with GLU=23%. However, a deeper and more quantitative analysis would require a detailed knowledge of several parameters (including e.g. the surface tension between the two phases, the size distribution of the droplets of dispersed phases, the composition and rheological properties of the continuous and dispersed phases) that are still unknown for the gluten samples. Interestingly, the complex viscosity of these samples is nearly constant in the range of frequencies investigated, with a value $\eta^* \sim 10$ Pa s. This numerical value is of the order of a pure gliadin sample at the same concentration prepared in an ethanol/water mixture which is a good solvent of these proteins [48]. However, our results are significantly different from those obtained by Kokini et al.[30], who measured aqueous gliadin-rich samples with no mention of phase separation and found essentially elastic samples (G'>G'') over a wide range of frequencies. The discrepancy could be due to the presence of large molecular weight glutenin polymers in their crude fraction. On the other hand, during a shear start-up experiment, we measure that the stress increases with time up to a maximum value, $\sigma_{Max}$, which grows with the applied shear rate, while the strain at the maximum stress, $\gamma_{Max}$, is constant. We find $\gamma_{Max} \cong 1$ (Fig. 5b), which may suggest that the stress transiently relaxes when the flow is strong enough to deform droplets by a distance equivalent to their size (100%). Interestingly, and uniquely for this first family of samples, the stress is relaxed through a pronounced undershoot before it reaches a steady state (Fig. 3a,b). The amplitude of the undershoot is larger and the time of its occurrence is faster as the shear rate increases (Fig. 6). Viscosity undershoots have been previously observed experimentally and numerically for a few polymeric systems [38, 49-53] and anisotropic colloid dispersions due to tumbling relaxation of anisotropic objects [40]. Interestingly, similar stress undershoots, in terms of amplitude and strain of occurrence, were also measured [39] and predicted in immiscible polymer melts [54]. These systems were treated as mixtures of two immiscible viscoelastic liquids whose interfacial area evolves under flow. The maximum stress was related to a slight deformation and orientation of the dispersed phase in the direction of the flow, while the stress undershoot was associated to an increase of the interfacial area due to the stretching of the dispersed phase [54]. In view of the biphasic nature of gliadin rich samples, their nonlinear behaviour could be rationalized considering a similar mechanism. Note that the increase of the amount of glutenin in the samples, from 7% to 23%, weakly affects the linear viscoelasticity but does not significantly impact the characteristics of the undershoot.

The second family corresponds to pre-gel samples, which are characterized by a balanced proportion of gliadin and glutenin (GLU=50 and 52%). Linear viscoelasticity measurements show G''>G' in the whole accessible frequency window, and tan $\delta$= G''/G'>1 decreases with frequency. However, a terminal regime, which would be characterized by G'~$\omega^2$ and G''~$\omega$ at low frequency is not reached, testifying for a non-fully relaxed state in the time scale probed experimentally. Accordingly, a shear-thinning behaviour is observed (Fig. 4), with a zero-shear viscosity not reached in the range of frequencies probed, giving evidence for their complex non-Newtonian behaviour. In addition, tan $\delta$ tends to reach a frequency-independent value at high frequency, as expected for a near-critical pre-gel state. Indeed, the critical state of a gel is reached when G'~ G''~ $\omega^\Delta$, and consequently tan $\delta$ is frequency-independent [21, 23]. The sample with GLU=52% is closer to its gel point than the sample with GLU=50%, as supported by higher shear moduli and a weaker evolution of tan $\delta$ with frequency. Microscopically, near critical pre-gel samples are characterized by power law distributions of cluster size (and hence of associated relaxation times) (Fig. 8), which shift towards higher values approaching the critical gel at percolation. Interestingly, in our experiments, very weak differences in terms of composition (GLU=50 and 52%) lead to a significant evolution of the viscoelastic properties of the two samples (Figs. 2, 4). This finding can be associated to a state very close to the critical gel for these two samples, where a divergence of viscosity is expected. On the other hand, shear start-up experiments reveal an overshoot of the transient viscosity η+ with a clear softening (η+ remaining below the linear envelope) before reaching a steady state (Fig. 3c,d). In polymeric systems, the occurrence of an overshoot is observed only when the applied shear rate is faster than the inverse of the terminal relaxation time. Our findings suggest that the terminal relaxation time is longer than 10s for the samples with GLU=50 and 52%. Furthermore, we find that the stress overshoot, $\sigma_{\text{Max}}$, is significantly higher for the sample closer to the critical gel point (GLU=52%) (Fig. 5a), whereas the maximum strain, $\gamma_{Max}$, is independent of the shear rate and the GLU content. Note though that $\gamma_{Max}$ is significantly higher for this second family ($\gamma_{Max} \approx 10$) as compared to the flowing samples of the first family ($\gamma_{Max} \approx 1$) (Fig. 5b). It indicates that the energy stored by the samples before their partial relaxation during shear increases with the GLU content due to the sample slower dynamics. It also corroborates with the ratio $\frac{\sigma_{\text{Max}}}{\sigma_{\text{Steady}}}$ significantly higher for the sample with GLU=52% than for the one with GLU=50% (Fig. 7), which indicates that, as the amount of GLU increases, clusters relax partially less stress, while being still deformed by the flow. Interestingly, high maximum strain values, such as those measured here for pre-gel samples ($\gamma_{Max} \cong 10$), are usually observed in branched polymeric systems [32, 55-57]. This is consistent with the structural view of near-critical pre gels [22] and percolation [58] on the one hand, and with recent experimental evidences for the branched structure of gluten pre-gels and gels with GLU=45% in water/ethanol solvent on the other hand [16]. This structural view of the pre-gel gluten samples is also consistent with the large size distribution of objects measured in the dilute regime using a fractionation technique that evidenced the joint presence of monomeric proteins, polymeric proteins and supramolecular assemblies with a branched structure when GLU ≥ 30% [31]. Finally, the good overlap of steady and complex viscosities within the experimental error (Fig. 3) validates the Cox-Merz rule [47], which discriminates any relevant irreversible change of samples due to the solicitation in the nonlinear regime.

The third family encompasses the two samples with the largest amounts of glutenin (GLU = 57% and GLU = 66%). These samples are in a gel state. Their linear viscoelasticity is characterized by G'>G'' (tan δ <1) with tan δ that increases with the frequency and moves toward a constant value at high frequency as expected for a near-critical post-gel state [21, 22]. Clearly, for these samples, the gel point is exceeded and a stress bearing network is formed, through the percolation of the clusters (Fig. 8). The structure is thought to form a permanent gel that does not relax at rest, as suggested by the fact that the elastic modulus tends to reach a plateau at low frequency. According to rubber elasticity theory, $G' \sim \xi^{-3}$, where $\xi$ is the network mesh size [59]. Assuming a same pre-factor for the two samples, we expect the mesh size to decrease by 20%, when GLU increases from 57 to 66% showing a clear impact of glutenin in building the elastic network. During shear start up tests, these samples strongly resist flow because of their solid-like viscoelasticity. Measurement issues are observed, such as gap opening and sample expulsion, as one forces the gels to flow. In future work, smaller geometries combined with cone and partitioned plate technology should be used for these samples to delay such instabilities, and more reliably characterize the strain hardening evidenced for the sample GLU = 57% at 0.1 s$^{-1}$ (Fig. 3e). The hardening is understood as the ability of the flow to induce a nonlinear stretching of the stress bearing chains due to the finite extensibility of the network strand [60-62]. The large value of the deformation at which strain-hardening is measured (around 10) is consistent with a loosely connected polymer-like gel sample.

Overall, the evolution of the gluten response to shear start-up, from pre-gel samples to post-gel samples is qualitatively similar to the response described for ionomers with different degrees of sulfonation: pre-gel samples show strain softening whereas strain hardening is measured for samples close to the gel point, and above the gel point samples fracture [63]. Shear start-up experiments on native gluten are scarce. Noticeable studies include the works described in references [11, 13] but which are limited to one shear rate and one sample composition. Interestingly these previous results are consistent with those obtained for the post-gel samples of the present study, in terms of strain hardening, maximum stress and maximum strain values. In addition, the authors also mentioned problems related to sample ejection from the rheometer. In native gluten the glutenin content (around 50%) is lower than in the present post-gel samples (GLU≥57%), and the fact that native gluten behaves as gel is presumably related to the presence of very high molecular weight glutenin polymers, which are insoluble in aqueous ethanol, and thus are absent in the model gluten extracts investigated in this work. Indeed, the extraction procedure used to extract model gluten with tunable composition discards these ethanol-water insoluble proteins [15]. Our unique protocol allows the formulation of well-controlled glutenin-rich samples and the detailed investigation of the role of glutenin in the mechanical and flow properties of gluten.

## Conclusion:

We have investigated the linear and nonlinear viscoelasticity of model gluten samples with a wide range of protein composition but a fixed total protein concentration (500g/L). Gliadin-rich samples form micro-phase separated dispersions in water whose rheological response is similar to that of viscoelastic emulsions. In particular, shear start-up experiments are characterized by the emergence of a stress undershoot possibly due to the elongation and breakup of droplets following their orientation in the flow. Modifying the protein composition and replacing part of the gliadins by glutenin polymers while keeping constant the total protein concentration enables to form monophasic transparent samples. We find that samples with an equal mass fraction of gliadins and glutenins (50 and 52% w/w glutenins) are in a viscoelastic pre-gel state and display shear-thinning and strain softening. Samples even more enriched in glutenins (57 and 66% w/w glutenins) are viscoelastic gels. Their mechanical response is similar to that of native gluten and is in particular characterized by strain hardening and sample instabilities in the shear flow. The linear and non-linear viscoelastic responses of the samples richer in glutenin than in gliadin suggest a microstructure akin to that of polymer near-critical gels made of polymer clusters with a power law distribution of size, which eventually percolate to yield a viscoelastic gel. Overall our findings show that glutenin polymers are responsible for the percolation of the gluten protein network and also facilitate the solubilization of gliadins in an aqueous solvent. The formation of gliadin-glutenin complexes soluble in water could explain this striking solubility evolution, but would require further investigation.

**Acknowledgments**

We acknowledge the French National Agency for funding of the project entitled Elastobio (ANR-18-CE06-0012-01).